\documentclass[nonacm,sigplan]{acmart}

\newcommand{\sys}{\textit{REACT}}
\newcommand{\mytilde}{\raise.17ex\hbox{$\scriptstyle\mathtt{\sim}$}}
\newcommand{\ie}{i.e.,}
\newcommand{\eg}{e.g.,}

\usepackage[normalem]{ulem}
\usepackage{pifont}

\usepackage{algorithm}
\usepackage{tabularx}
\usepackage{mathtools}
\usepackage{subfig}
\usepackage{colortbl}
\usepackage{multirow}
\usepackage[normalem]{ulem}

\usepackage[font=small,labelfont=bf]{caption}

\newcommand{\hrw}[1]{}

\begin{document}

\title{Energy-adaptive Buffering for Efficient, Responsive, and Persistent Batteryless Systems}

\author{Harrison Williams}
\affiliation{%
    \institution{Virginia Tech}
    \country{}
}
\email{hrwill@vt.edu}

\author{Matthew Hicks}
\affiliation{%
    \institution{Virginia Tech}
    \country{}
}
\email{mdhicks2@vt.edu}

\begin{abstract}
Batteryless energy harvesting systems enable a wide array of new sensing, computation, and communication platforms untethered by power delivery or battery maintenance demands.
Energy harvesters charge a buffer capacitor from an unreliable environmental source until enough energy is stored to guarantee a burst of operation despite changes in power input.
Current platforms use a fixed-size buffer chosen at design time to meet constraints on charge time or application longevity, but static energy buffers are a poor fit for the highly volatile power sources found in real-world deployments: fixed buffers waste energy both as heat when they reach capacity during a power surplus and as leakage when they fail to charge the system during a power deficit.

To maximize batteryless system performance in the face of highly dynamic input power, we propose \sys{}: a responsive buffering circuit which varies total capacitance according to net input power.
\sys{} uses a variable capacitor bank to expand capacitance to capture incoming energy during a power surplus and reconfigures internal capacitors to reclaim additional energy from each capacitor as power input falls.
Compared to fixed-capacity systems, \sys{} captures more energy, maximizes usable energy, and efficiently decouples system voltage from stored charge---enabling low-power and high-performance designs previously limited by ambient power.
Our evaluation on real-world platforms shows that \sys{} eliminates the tradeoff between responsiveness, efficiency, and longevity, increasing the energy available for useful work by an average 25.6\% over static buffers optimized for reactivity and capacity, improving event responsiveness by an average $7.7x$ without sacrificing capacity, and enabling programmer directed longevity guarantees.
\end{abstract}

\maketitle 
\pagestyle{plain} 

\section{Introduction}
\label{sec:introduction}
Ever-shrinking computing and sensing hardware has pushed mobile Internet-of-Things (IoT) type devices beyond the limits of the batteries powering them.
A typical low cost/power microcontroller~\cite{msp430g2252} drains a 1 $cm^3$ battery nearly $14$x its size in just over 8 weeks of active operation~\cite{cr2032}, rendering the system useless without a potentially costly replacement effort.
Cost, maintenance, and safety concerns make batteries further incompatible with massive-scale (one million devices per square kilometer~\cite{what-is-mmtc}) and deeply-deployed (infrastructure~\cite{concrete}, healthcare~\cite{batteryless-ingestible}) applications.
IoT engineers are turning to batteryless energy harvesting platforms to power low-cost, perpetual systems capable of driving a ubiquitous computing revolution.
Increasingly efficient energy harvesting circuits enable batteryless systems across a range of IoT use cases including feature-rich batteryless temperature sensors $500$x smaller than a grain of rice~\cite{rice-computer} and batteryless flow-meters~\cite{flowmeter} supporting deep-sea drilling or geothermal plants for decades without maintenance.

The energy harvesting design model both enables new deployments previously limited by batteries and places new demands on system developers.
Harvested energy is highly unreliable: sensitive environmental factors such as shadows over a photovoltaic cell or shifts in the orientation of a rectenna produce rapid, outsized changes in the energy scavenged by the harvester.
Energy harvesters mitigate this unreliability by charging an energy buffer to a given enable voltage, which the system periodically discharges to supply a useful quantum of work despite potential power loss.

Buffer capacity is a key design element of any batteryless system.
Past work~\cite{capybara} explores the tradeoff between buffer sizes: small buffers are highly \textit{reactive}---charging rapidly and quickly enabling the system to address time-sensitive events---but sacrifice \textit{longevity} because they rapidly discharge during operation, guaranteeing only a short burst of uninterrupted operation.
Large buffers store more energy at a given voltage, improving longevity by supporting a longer or more energy-intensive burst of operation at the cost of reactivity because they require more energy to enable the system at all.
Matching buffer size to projected energy demand is critical to ensuring the system is both reactive enough to address incoming events/deadlines (\eg{} periodic sensor readings) and long-lived enough to support uninterruptible operations (\eg{} radio transmissions).
Designers choose the minimum size necessary to power all atomic operations on the device, maximizing reactivity given a required level of longevity.

In this work, we explore static energy buffer \textit{efficiency} as a third metric for buffer performance and find that it varies dramatically with \textit{net energy input} rather than simple energy demand.
Small buffers reach capacity quickly if power input exceeds instantaneous demand---burning off hard-won energy as heat to prevent overvoltage.
Large buffers capture all incoming power, but enable slowly and lose more harvested energy to leakage below the minimum system voltage.
The volatile nature of harvested power means that fixed-size buffers experience \textit{both} problems over the course of their deployment, discharging energy during a power surplus and losing large portions of energy to leakage during a deficit.

To make the most of incoming energy in all circumstances, we propose \sys{}\footnote{\textbf{R}econfigurable, \textbf{E}nergy-\textbf{A}daptive \textbf{C}apaci\textbf{T}ors}: a dynamic energy buffering system that varies its capacitance following changes in \textit{net} power.
\sys{} maximizes system responsiveness and efficiency using a small static buffer capacitor, quickly enabling the system to monitor events or do other low-power work under even low input power.
If input power rises beyond the current system demand and the static buffer approaches capacity, \sys{} connects additional capacitor banks to absorb the surplus, yielding the capacity benefits of large buffers without the responsiveness penalty.
When net power is negative, these capacitors hold the system voltage up and extend operation beyond what is possible using the small static capacitor.

While expanding buffer size to follow net power input ensures the system can capture \textit{all} incoming energy, increasing capacitance also increases the amount of unusable charge stored on the capacitor banks---charge which could power useful work if it were on a smaller capacitor and therefore available at a higher voltage.
As supply voltage falls and approaches a minimum threshold, \sys{} \textit{reclaims} this otherwise-unavailable energy by reconfiguring capacitor banks into series, shifting the same amount of charge onto a smaller equivalent capacitance in order to boost the voltage at the buffer output and ensure the system continues operating for as long as possible.
\sys{} effectively eliminates the design tradeoff between reactivity and capacity by tuning buffer size within an arbitrarily large capacitance range, only adding capacity when the buffer is already near full.
\sys{}'s charge reclamation techniques maximize efficiency by moving charge out of large capacitor banks onto smaller ones when net input power is negative, ensuring all energy is available for useful work.


We integrate a hardware prototype of \sys{} into a full energy harvesting platform to evaluate it against previous work, operating under different input power conditions and with different power consumption profiles.
Our results indicate that \sys{} provides the "best of both worlds" of both small- and large-buffer systems, rapidly reaching the operational voltage under any power conditions while also expanding as necessary to capture all available energy and provide software longevity guarantees as needed.
Maximizing buffer capacity and reclaiming charge using \sys{}'s reconfigurable capacitor banks eliminates the efficiency penalties associated with both small and large static capacitor buffers, increasing the portion of harvested energy used for application code by an average 39\% over an equally-reactive static buffer and 19\% over an equal-capacity one.
Compared to prior work exploring dynamic capacitance for batteryless systems~\cite{morphy}, \sys{} improves performance by an average 26\% owing to its efficient charge management structure.
This paper makes the following technical contributions:
\begin{itemize}
    \item We evaluate the power dynamics of common batteryless systems in real deployments and explore how common-case volatility introduces significant energy waste in static or demand-driven buffers (\S{}~\ref{sec:background}).
    \item We design \sys{}, a dynamic buffer system which varies its capacitance according to system needs driven by net input power (\S{}~\ref{sec:design}). \sys{}'s configurable arrays combine the responsiveness of small buffers with the longevity and capacity of large ones, enables energy reclamation to make the most of harvested power, and avoids the pitfalls of energy waste inherent in other dynamic capacitance designs (\S{}~\ref{sec:capacitor-banks}).
    \item We integrate \sys{} into a batteryless system and evaluate its effect on reactivity, longevity, and efficiency under a variety of power conditions and workloads (\S{}~\ref{sec:evaluation}). Our evaluation indicates that \sys{} eliminates the responsiveness-longevity tradeoff inherent in static buffer design while increasing overall system efficiency compared to \textit{any} static system.
\end{itemize}


\section{Background and Related Work}
\label{sec:background}
Scaling sensing, computation, and communication down to smart dust~\cite{dust} dimensions requires harvesting power on-demand rather than packaging energy with each device using a battery.
Many batteryless systems use photovoltaic~\cite{rice-computer, gameboy} or RF power~\cite{mementos}, while other use-cases are better suited for sources such as vibration~\cite{piezo}, fluid flow~\cite{flowmeter}, or heat gradients~\cite{nvprocessor}.
Commonalities between ambient power sources have inspired researchers to develop general-purpose batteryless systems, regardless of the actual power source: ambient power is \textit{unpredictable}, \textit{dynamic}, and often \textit{scarce} relative to the active power consumption of the system.

Batteryless systems isolate sensing and actuation components from volatile power using a buffer capacitor.
The harvester charges the capacitor to a pre-defined enable voltage, after which the system turns on and begins consuming power.
Because many environmental sources cannot consistently power continuous execution, systems operate intermittently---draining the capacitor in short bursts of operation punctuated by long recharge periods.
This general-purpose intermittent operation model has enabled researchers to abstract away the behavior of incoming power and focus on developing correct and efficient techniques for working under intermittent power~\cite{dino, ratchet, clank, chinchilla, totalrecall, alpaca}.

\subsection{Choosing Buffer Capacity}
\label{sec:choosing-capacity}
\begin{figure}[t]
  \centering
  \includegraphics[width=\columnwidth]{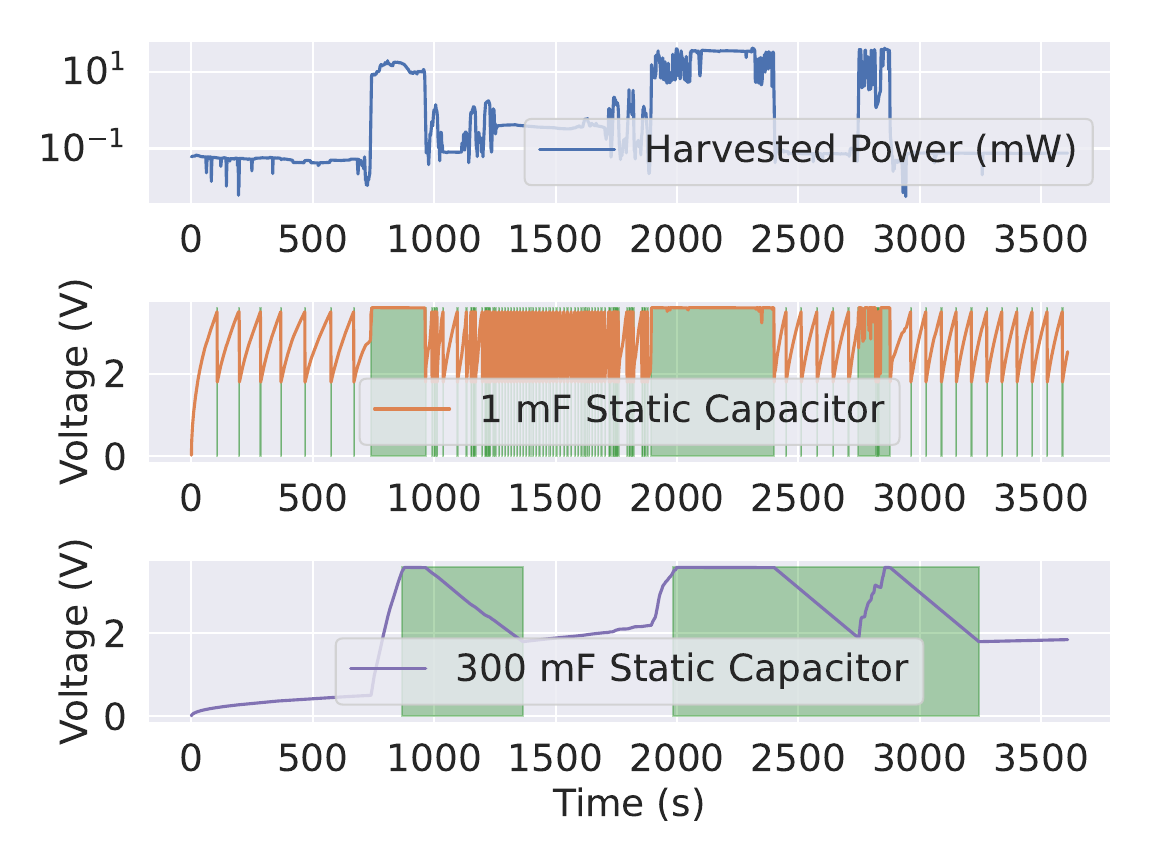}
  \caption{Static buffer operation on a simulated solar harvester. Highlighted blocks indicate the system is running.}
  \label{fig:solar_comparison}
  \hrulefill
\end{figure}
Buffer size determines system behavior in several important ways.
Supercapacitors provide inexpensive and small-form-factor bulk capacitance~\cite{supercap}, enabling designers to choose a capacitor according to performance rather than cost or size concerns.
Two metrics motivate past work: \textit{reactivity} refers to the system's ability to rapidly charge to its enable voltage and begin operation.
High reactivity ensures a system is online to execute periodic tasks or address unpredictable input events.
\textit{Longevity} refers to the energy available for an uninterrupted period of work with no additional power input; long-lived systems support high-power and long-running uninterruptible operations and reduce the overhead incurred with state recovery after a power loss.

\subsubsection{Reactivity and Longevity:}
A batteryless system's reactivity and longevity depend primarily on the charge and discharge rate of the buffer capacitor.
We illustrate the tradeoff using a simulated solar harvester with a 22\% efficient, 5 cm\textsuperscript{2} panel, based on a pedestrian trace from the EnHANTs solar dataset~\cite{enhants}.
The system runs from 3.6V down to 1.8V and draws 1.5 mA in active mode, representative of a typical deployment~\cite{efm32wg}.
Figure~\ref{fig:solar_comparison} illustrates the reactivity-longevity tradeoff inherent in static buffer systems at two design extremes, using a 1 mF and 300 mF capacitor.
The 1 mF system charges rapidly and is therefore highly reactive, reaching the enable voltage over $8x$ sooner than the 300 mF version.
However, the smaller capacitor also discharges quickly---the mean length of an uninterrupted power cycle using the 1 mF capacitor is 10 seconds versus 880 seconds for the 300 mF capacitor, indicating the 300 mF system is far longer-lived once charged.
The relative importance of reactivity and longevity depends on the use case, but often changes over time for a complex system---complicating design further.

\subsubsection{Power Volatility and Energy Efficiency:}
Buffer capacity is also a major driver of end-to-end energy \textit{efficiency}: using the 300 mF capacitor our system is operational for 49\% of the overall power trace, compared to only 27\% for the 1 mF platform.
This stems from the high volatility of incoming power---82\% of the total energy input is collected during short-duration power spikes when harvested power rises above 10 mW, despite the system spending 77\% of its time at input powers below 3 mW.
A large buffer captures this excess energy to use later while the smaller buffer quickly reaches capacity and discharges energy as heat to avoid overvoltage.

Large buffers, however, are not always more efficient: the energy used to charge the capacitor to the operational voltage cannot power useful work, and is eventually lost to leakage while the system is off.
When power input is low, this "cold-start" energy represents a significant portion of total harvested energy.
For the system described above powered by a solar panel at night~\cite{enhants}, the 1 mF buffer enables a duty cycle of 5.7\% versus only 3.3\% using a 10 mF buffer.
This low power environment highlights another risk of oversized buffers: the system using the 300 mF capacitor never reaches the enable voltage and so never begins operation.

Improvements in harvester efficiency and low-power chip design are closing the gap between harvester output and chip power consumption.
Power is increasingly limited by volatile environmental factors rather than scarcity induced by low efficiency; the result is that energy harvesters experience periods of both energy scarcity and surplus.
Rapidly changing power conditions place opposing demands on batteryless systems, which must remain responsive with low input power, provide longevity for long-running operations, and maximize efficiency by avoiding energy waste.

\subsection{Power-Responsive Performance Scaling}
One solution to volatile energy input is modulating throughput according to incoming power, increasing execution rate when power is plentiful and decreasing it to maintain availability when power is scarce.
Limiting net input power to the buffer by matching power consumption with input enables systems to use small buffer capacitors without reaching the buffer capacity, ensuring no power is wasted with an over-full buffer.
Past work realizes power-responsive scaling using heterogeneous architectures~\cite{phase} or by adapting the rate and accuracy of software execution~\cite{rehash, catnap, astar, d2vfs}.

Unfortunately, we find the assumptions underlying power-performance scaling often do not apply to batteryless systems.
Increasing energy consumption by accelerating execution only improves systems which \textit{have useful work to do exactly when input power is high}, but many batteryless systems focus on periodic sensing and actuation deadlines which do not correlate with ambient power supply.
Further, resource-constrained platforms may have few on-chip operations which can be delayed until power is plentiful; when these operations do exist, they are often not amenable to scaling (\eg{} transmitting data to a base station may be delayed but always requires a fixed-cost radio operation).
Flexible batteryless systems must capture energy and use it on demand rather than fit operation to unreliable power input.

\subsection{Multiplexed Energy Storage}
Rather than match power consumption to incoming supply, systems may charge multiple static buffers according to projected demand.
Capybara~\cite{capybara} switches capacitance using an array of heterogeneous buffers: programmers set capacitance modes throughout the program, using a smaller capacitor to maximize responsiveness for low-power or interruptible tasks and switching to a larger capacitor for high-power atomic operations.
UFoP and Flicker~\cite{federation, flicker} assign each peripheral on the system a separate buffer and charging priority, enabling responsive low-power operation while waiting to collect sufficient energy for high-power tasks.
These systems increase overall energy capacity by directing excess power to capacitors not currently in use.

Static arrays increase capacity without reducing responsiveness, but waste energy when charge is stored on unused buffers.
Reserving energy in secondary capacitors 1) requires error-prone~\cite{catnap} speculation about future energy supply and demand to decide charging priorities, which can change between when energy is harvested and when it needs to be used; and 2), wastes energy as leakage when secondary buffers are only partially charged, failing to enable associated systems and keeping energy from higher-priority work.
To minimize programmer speculation, decouple tasks which compete for buffered energy, and minimize leakage, energy must be \textit{fungible}: the buffer must be capable of directing \textit{all} harvested energy to \textit{any} part of the system on demand.

\subsection{Unified Dynamic Buffering}
Past work has also explored varying the behavior of a single unified buffer to capture the fungibility requirement described above.
Dewdrop~\cite{dewdrop} varies the enable voltage to draw from a single capacitor according to projected needs (\eg{} begin operation at 2.2V instead of 3.6V)---providing complete energy fungibility---but still suffers from the reactivity-longevity tradeoff of capacitor size.
Morphy~\cite{morphy} replaces static buffers using a set of capacitors in a unified switching network; software can connect and disconnect arbitrary sets of capacitors in series or parallel to produce different equivalent capacitances.
Morphy addresses two challenges facing batteryless systems: fast charging from low-voltage sources by operating the capacitor network as a charge pump, and charge isolation to allocate set amounts of energy to tasks and prevent buggy code from discharging the entire buffer by isolating parts of the capacitor array.

We evaluate \sys{} alongside Morphy because the Morphy architecture can also be used to target the reactivity and longevity challenges discussed in \S{}~\ref{sec:introduction}.
By starting with a small equivalent capacitance, the system rapidly reaches an operational voltage.
If the buffer as configured reaches capacity, software can gradually reconfigure the capacitor network to increase capacitance.
Charge flows between the capacitors to equalize the voltage on the capacitor array and reduce the output voltage of the network, enabling the bank to harvest more energy without reaching capacity.
However, this current flow between capacitors to equalize voltage during reconfiguration dissipates a significant amount of the energy stored in the network (we explore these energy dynamics in detail in \S{}~\ref{sec:capacitor-organization}).
Our evaluation in \S{}~\ref{sec:evaluation} shows that this internal power dissipation reduces common-case end-to-end performance to below that of systems using appropriately-sized static capacitors, making this approach impractical for energy-constrained devices.
An energy-focused approach prioritizing minimal power dissipation is key to developing intermittent systems that can simultaneously maximize reactivity, longevity, and overall efficiency.

\section{Design}
\label{sec:design}
An intelligent energy buffering strategy is key to effective and efficient batteryless systems.
Three performance objectives, informed by the advantages and limitations of prior approaches, drive \sys{}'s design:
\begin{itemize}
    \item \textbf{Minimize charge time:} Rapidly reaching the operational voltage, even when buffered energy cannot support complex operation, maximizes reactivity and enables systems to reason about power or sensor events from within low-power sleep modes.
    \item \textbf{Maximize capacity:} System-wide longevity and efficiency require buffering large amounts of incoming energy when power supply exceeds demand, either to power long-running uninterruptible operations or support future power demand when supply is low.
    \item \textbf{Maximize energy fungibility:} Unpredictable power demand patterns mean that energy cannot be pre-provisioned to specific operations at harvest-time; systems need the ability to draw all harvested energy from the buffer and direct it as needed.
\end{itemize}


\subsection{\sys{} Overview}
\begin{figure}[t]
  \centering
  \includegraphics[width=0.85\columnwidth]{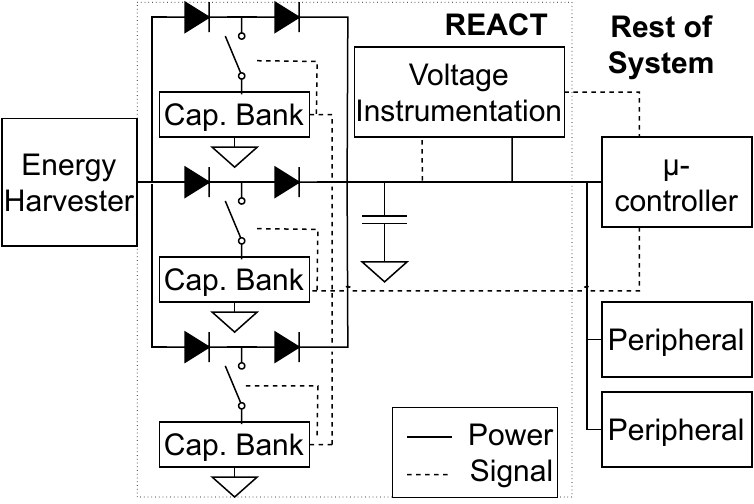}
  \caption{\sys{} diagram and signal flow between components.}
  \label{fig:block_diagram}
  \hrulefill
\end{figure}
\sys{} buffers energy using a fabric of reconfigurable capacitor banks that software adjusts as needed.
Figure~\ref{fig:block_diagram} shows a high-level overview of \sys{}'s hardware design.
We design \sys{}'s hardware component as a drop-in replacement for a typical buffer between the harvester and the rest of the system, while the buffer management software requires no code modification or programmer input.
The only system prerequisite is a set of digital I/O pins to configure capacitor banks and receive voltage monitoring information.


\subsection{Cold-start Operation and the Last-level Buffer}
From a cold start ($E(t)=0$), \sys{} minimizes overall capacitance in order to rapidly charge to the operational voltage and enable the system with minimum energy input (high reactivity).
The minimum capacitance is set by the smallest quantum of useful work available on the system (minimum required longevity), such as a short-lived software operation or an initialization routine that puts the system into a low-power responsive sleep mode.
\sys{} provides this rapid charge time using a small static buffer referred to hereafter as the \textit{last-level buffer}.
Additional capacitor banks are connected using normally-open switches and only contribute to overall capacitance when configured to do so in software, after the system is able to reason about buffered energy.

The last-level buffer sets the minimum capacitance at power-on when all other banks are disconnected.
This enables simple tuning of the energy input required to enable the system (reactivity) and the guaranteed energy level when the system does begin work (minimum longevity).
It also smooths voltage fluctuations induced by capacitor bank switching (\S{}~\ref{sec:charge-reclamation}).
Finally, the last-level buffer serves as the combination point between the different capacitor banks and the rest of the system.
Although energy may be stored in multiple banks of varying capacity at different voltages, combining it at the last-level buffer simplifies system design by presenting harvested power as a unified pool of energy which the system taps as needed (\ie{} harvested energy is fungible).

\subsubsection{Monitoring Buffered Energy}
Despite mutual isolation, bank voltages tends to equalize: the last-level buffer pulls energy from the highest-voltage bank first, and current flows from the harvester to the lowest-voltage bank first.
This enables \sys{} to measure only the voltage on the last-level buffer as a surrogate for remaining energy capacity.
If voltage rises beyond an upper threshold---the buffer is near capacity---\sys{}'s voltage instrumentation hardware signals the software component running on the microcontroller to increase capacitance using the configurable banks.
Voltage falling below a lower threshold indicates the buffer is running out of energy and that \sys{} should reconfigure banks to extract additional energy and extend operation.
\sys{}'s instrumentation only needs to signal three discrete states---near capacity, near undervoltage, and OK---so two low-power comparators is sufficient for energy estimation.

\subsection{Dynamic Capacitor Banks}
\label{sec:capacitor-banks}
\begin{figure}[t]
  \centering
  \subfloat[\centering{}\footnotesize{}Two capacitors, parallel configuration\label{fig:bank_diagram_2}]{%
    \includegraphics[width=0.35\columnwidth]{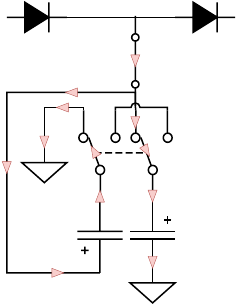}
  }\hspace{0.1\columnwidth}
  \subfloat[\centering{}\footnotesize{}Three capacitors, series configuration\label{fig:bank_diagram_3}]{%
    \includegraphics[width=0.35\columnwidth]{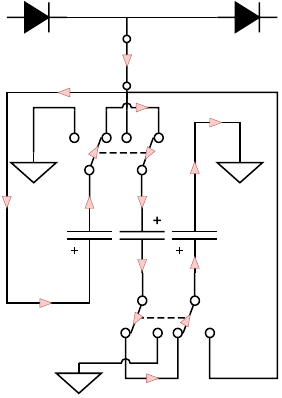}
  }
  \caption{\sys{} capacitor banks in different bank sizes and configurations. Arrows indicate charging current path.}
  \label{fig:bank_diagrams}
  \hrulefill
\end{figure}
The last-level buffer on its own enables high reactivity and minimizes cold-start energy below the operational minimum, maximizing efficiency during power starvation.
However, when net power into the buffer is positive---such as during a period of high input power or low workload---the small last-level buffer rapidly reaches capacity.
\sys{} provides the energy capacity required to both maximize efficiency and support long-running operation by connecting configurable capacitor banks when the last-level buffer reaches capacity, as shown in Figure~\ref{fig:block_diagram}.

\subsubsection{Capacitor Organization}
\label{sec:capacitor-organization}
\begin{figure}[t]
  \centering
  \subfloat[\centering{}\footnotesize{}Full series configuration with $C_{eq}$ = C/4\label{fig:morphy-series}]{%
    \includegraphics[width=\columnwidth]{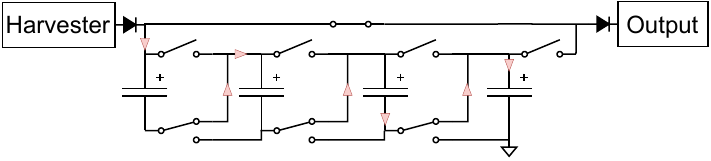}
  }
  \vspace{-10pt}
  \subfloat[\centering{}\footnotesize{}Full parallel configuration with $C_{eq}$ = 4C\label{fig:morphy-parallel}]{%
    \includegraphics[width=\columnwidth]{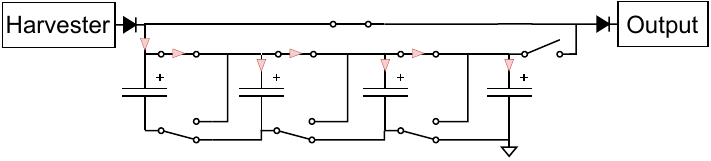}
  }
  \caption{Structure of the unified approach presented by Yang et al.~\cite{morphy}. Arrows indicate charging current path.}
  \label{fig:morphy-intro}
  \hrulefill
\end{figure}
\begin{figure}[t]
\centering
  \includegraphics[width=0.8\columnwidth]{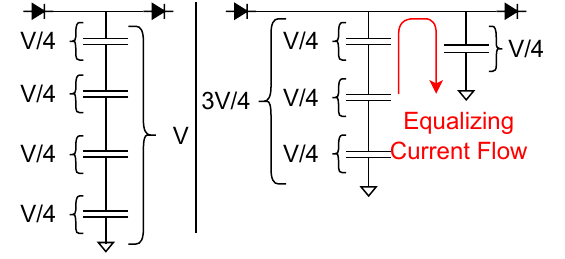}
  \caption{Dissipative current flow in a fully-unified buffer during reconfiguration. Energy is dissipated by current spikes after capacitors at different voltages are placed in parallel.}
  \label{fig:morphy-problem}
  \hrulefill
\end{figure}
Careful management of the connections between each capacitor is key to maximizing energy efficiency while also presenting a valid operational voltage for the computational backend.
Morphy~\cite{morphy} presents one approach: by connecting a set of equally-sized capacitors through switches similar to a charge pump, overall buffer capacitance can be varied across a wide range of capacitance values.
Different switch configurations produce intermediate buffer sizes between the extremes shown in Figure~\ref{fig:morphy-intro}; gradually stepping through these configurations smoothly varies capacitance through software control.

A fully interconnected array enables a wide range of equivalent, but introduces significant waste through dissipative heating when the charged capacitor array is reconfigured.
Figure~\ref{fig:morphy-problem} illustrates how energy is lost when charged capacitors are connected in a new configuration.
Before reconfiguration, the energy contained in the system is $E_{old}=\frac{1}{2}(C/4)V^2$; when a capacitor is taken out of series and placed in parallel with the remaining capacitors to increase equivalent capacitance to $4C/3$, current flows to the lower-voltage newly-parallel capacitor to equalize output voltage.
The final output voltage is $3V/8$, and the remaining energy is $E_{new}=\frac{1}{2}(4C/3)(3V/8)^2$.
The portion of energy conserved is $E_{new}/E_{old} = 0.75$---\ie{} 25\% of buffered energy is dissipated by current in the switches during reconfiguration.
Larger arrays are increasingly inefficient: the same scenario with an 8-capacitor array wastes 56.25\% of its buffered energy transitioning from an 8-parallel to a 7-series-1-parallel configuration.
Similar waste occurs when reducing equivalent capacitance by placing capacitors in series.\footnote{Charge pumps avoid this waste by never connecting capacitors at different potentials in parallel; in this use case, however, parallel capacitance is always necessary to smooth voltage fluctuations during switching and keep the output voltage within the computational backend's acceptable range.}
Our evaluation in \S{}~\ref{sec:performance} indicates that the energy loss caused by switching often outweighs any advantage from dynamic behavior, causing the fully-connected approach to underperform even static buffers.

\subsubsection{Bank Isolation}
The switching loss discussed above stems from charge flowing between capacitors within the power network as they switch into different configurations.
\sys{} eliminates unnecessary current flow by organizing capacitors into independent, mutually isolated banks as shown in Figure~\ref{fig:block_diagram}.
Figure~\ref{fig:bank_diagrams} illustrates in detail two example capacitor banks in each possible configuration: capacitors within a bank can only be arranged in either full-series (low capacitance) or full-parallel (high capacitance) so that no current flows between capacitors \textit{within} a bank.
Isolation diodes on the input and output of each bank prevent current \textit{between} banks: when a charged parallel-configured bank is reconfigured into series (reducing its capacitance and boosting its output voltage), isolation diodes prevent it from charging other banks in the array.
Similarly, banks re-configured into parallel cannot draw current from anywhere except the energy harvester.
Isolation reduces the number of potential capacitor configurations compared to a fully-connected network, but dramatically increases energy efficiency.

\sys{}'s isolation diodes direct the flow of current: intermediate capacitor arrays are only charged directly from the energy harvester and only discharge to the last-level buffer.
This also means that all current from the harvester flows through two diodes before reaching the system, so minimizing power dissipation in the diodes is essential to maintaining overall system efficiency.
To maximize charging efficiency, we design \sys{} using ideal diode circuits incorporating a comparator and pass transistor, rather than typical PN or Schottky diodes.
Active ideal diodes are far more efficient at typical currents for batteryless systems: the circuit we use~\cite{lm66100} dissipates $0.02\%$ of the power dissipated in a typical Schottky diode~\cite{schottky} at a supply current of 1 mA.

\subsubsection{Bank Reconfiguration}
\label{sec:reconfiguration}
The range of buffer sizes depends on the number of capacitor banks and the number of capacitors in each bank.
\sys{}'s capacitor banks are effectively connected in parallel, so the overall capacitance is the sum of each bank's contribution.
Each \sys{} bank containing $N$ identical capacitors of capacitance $C$ may be configured to contribute no capacitance (disconnected), series capacitance $C/N$, or parallel capacitance $NC$.

\sys{} must increment buffer capacitance in small steps in order to keep voltage within the operational range while capturing all incoming power.
A large increase in capacitance pulls output voltage down and introduces cold-start energy loss if net power input is low; for extreme cases, the system may run out of energy and cease execution while the new capacitance charges \textit{even if incoming power would be sufficient to power operation}.
\sys{} first connects banks in the series configuration to contribute a small capacitance and avoid large jumps in overall buffer size.
If the buffer continues to charge and reaches the upper voltage limit $V_{high}$, \sys{} further expands capacitance by toggling double-pole-double-throw bank switches to configure the capacitors in parallel.
Expanding the buffer by reconfiguring charged capacitors rather than adding new ones reduces the time the system is cut off from input power while current flows exclusively to the new capacitance, because it is already charged to $V_{high}/N$.
Because no current flows between capacitors or banks, bank reconfiguration changes capacitance seen on the common rail without dissipative loss.
\sys{} uses break-before-make switches to ensure no short-circuit current flows during switching; incoming current flows directly to the last-level buffer during the momentary open-circuit in the bank.

\subsubsection{Charge Reclamation}
\label{sec:charge-reclamation}
Reconfiguring a bank from series to parallel allows \sys{} to efficiently increase capacitance without dropping output voltage.
When voltage on the last-level buffer appraches the threshold value $V_{low}$, indicating net power is leaving the buffer, \sys{} needs to reduce equivalent capacitance to boost voltage and keep the backend running.
\sys{} accomplishes this by transitioning charged $N$-capacitor banks from the parallel to the series configuration, reducing equivalent capacitance from $NC$ to $C/N$ and boosting output voltage from $V_{low}$ to $NV_{low}$.
This boosts voltage on the last-level buffer and extracts more energy from the capacitor bank than would otherwise be available once voltage falls below $V_{low}$.

The remaining energy unavailable after the parallel→series transition depends on the number $N$ of $C_{unit}$-size capacitors in the bank.
Before switching, the cold-start energy stored on the parallel-mode bank is $E_{par} = \frac{1}{2}NC_{unit}V_{low}^2$.
Switching to the series configuration conserves stored energy: $E_{ser}=\frac{1}{2}(C_{unit}/N)(NV_{low})^2=E_{par}$, but boosts voltage to enable the digital system to continue extracting energy.
If net power remains negative, the system eventually drains the series-configuration bank down to $V_{low}$.
This is energetically equivalent to draining the parallel-configuration bank to $V_{low}/N$, leaving $E_{par}=\frac{1}{2}NC_{unit}(V_{low}/N)^2=\frac{1}{2}C_{unit}V_{low}^2/N$ unusable; the overall result is that \sys{} reduces energy loss by a factor of $N^2$ when reducing system capacitance compared to simply disconnecting the capacitor.

\subsubsection{Bank Size Constraints}
\label{sec:size-constraints}
Increasing the number of capacitors $N$ in a bank improves efficiency by reclaiming more energy when switching a bank from parallel to series.
However, it also introduces voltage spikes when the bank output voltage is temporarily multiplied by $N$, increasing dissipative loss as current flows from the bank to the last-level buffer.
Because \sys{} measures overall energy at the last-level buffer, the software component may interpret this voltage spike as a buffer-full signal and incorrectly add capacitance despite low buffered energy.
In extreme cases, the voltage spike may exceed component absolute limits.

The size of the last-level buffer $C_{last}$ constrains the number $N$ and size $C_{unit}$ of each capacitor in a bank in order to keep voltage below \sys{}'s buffer-full threshold during a parallel→series transition.
A larger $C_{unit}$ contains more energy and thus pulls voltage higher when switched from parallel to series.
Equation~\ref{eq:final-voltage} gives the last-level buffer voltage after switching a bank to series at a trigger voltage $V_{low}$:
\small
\begin{equation}
    \label{eq:final-voltage}
    V_{new} = \frac{(NV_{low})(C_{unit}/N)}{C_{last}+C_{unit}/N} + \frac{V_{low}*C_{last}}{C_{last}+C_{unit}/N}
\end{equation}
\normalsize
Constraining $V_{new}<V_{high}$ and solving for $C_{unit}$ yields the absolute limit for $C_{unit}$ (Equation~\ref{eq:unit-limit}).
Note that $C_{unit}$ is only constrained if the parallel→series transition at $V_{low}$ produces a voltage above $V_{high}$:
\small
\begin{equation}
    \label{eq:unit-limit}
    C_{unit} < \frac{NC_{last}(V_{high}-V_{low})}{NV_{low}-V_{high}}
\end{equation}
\normalsize

\subsection{\sys{} Software Interface}
\sys{}'s standalone hardware design means that the software component running on the target microcontroller is minimal.
The software subsystem monitors for incoming over- or under-voltage signals from \sys{}'s voltage instrumentation and maintains a state machine for each capacitor bank.
Each capacitor bank is disconnected at startup; on an overvoltage signal from \sys{}'s hardware, the software directs \sys{} to connect a new capacitor bank in the series configuration.
A second overvoltage signal\footnote{\sys{} polls the over/undervoltage signals using an internal timer rather than edge-sensitive interrupts to handle cases such as a high enough power input that the capacitance step does not pull supply voltage below $V_{low}$.} causes \sys{} to reconfigure the newly-connected bank to parallel; on the next overvoltage signal, \sys{} connects a second capacitor bank, and so on.
\sys{} similarly steps capacitor banks in the opposite direction when an undervoltage signal arrives.

\subsubsection{Software-Directed Longevity}
\label{sec:software-longevity}
\sys{}'s software component requires no active programmer intervention or code changes aside from setting voltage thresholds, initializing each bank state machine, and setting the order to connect and disconnect banks.
Software does not need to know the details ($N$, $C_{unit}$) of each bank, although this information with the state of each bank gives a coarse idea of the current buffered energy.
Because \sys{} only changes capacitance when the bank is near-full or near-empty, capacitance level is an effective surrogate for stored energy.
Application code can use this feedback to set longevity guarantees through \sys{}'s software interface.
In preparation for a long-running or high-energy atomic operation, software sets a \textit{minimum capacitance level} corresponding to the amount of energy required and then enters a deep-sleep mode keeping \sys{}'s capacitor polling time active.
As the system charges, \sys{} eventually accumulates enough energy to reach the minimum capacitance level---indicating that enough energy is stored to complete the planned operation, and pulling the system out of its deep-sleep with enough energy to complete execution regardless of future power conditions.

\section{Implementation}
\label{sec:implementation}
\begin{table}[t]
\resizebox{\columnwidth}{!}{
\centering
\footnotesize{}
\begin{tabular}{|c|c|c|c|c|c|c|}
\hline
\textbf{Bank}       & \textbf{0} & \textbf{1} & \textbf{2} & \textbf{3} & \textbf{4} & \textbf{5} \\ \hline
\textbf{Capacitor Size ($\mu{}F$)}  & 770       & 220       & 440       & 880      & 880       & 5000      \\ \hline
\textbf{Capacitor Count} & 1          & 3          & 3          & 3          & 3          & 2          \\ \hline
\end{tabular}
}
\caption{Bank size and configurations for our \sys{} test implementation. Bank  0 is the last-level buffer.}
\label{table:caps}
\hrulefill
\end{table}
We explore \sys{}'s impact on overall efficiency, reactivity, and longevity using a hardware prototype integrated into a real batteryless platform.
All files for \sys{}, the baseline systems, and the energy harvesting emulator will be open-sourced upon publication.
Our testbed is based on the MSP430FR5994~\cite{fr5994}, a popular microcontroller for energy harvesters~\cite{failureSentinels, totalrecall, alpaca}.
For each buffer configuration we evaluate, an intermediate circuit power gates the MSP430 to begin operation once the buffer is charged to 3.3V and disconnects it when the buffer voltage reaches 1.8V.

Our \sys{} implementation has a range of 770 $\mu{}F$-18.03 $mF$ using a set of 5 dynamic banks, in addition to the last-level buffer, detailed in Table~\ref{table:caps}.
We implement the capacitors in banks 0-4 using combinations of 220 $\mu{}F$ capacitors with max leakage current of 28 $\mu{}A$ at their rated voltage of 6.3V~\cite{myCeramics}.
Bank 5 uses supercapacitors with approximately 0.15 $\mu{}A$ leakage current at 5.5V~\cite{mySupercap}.

\subsection{Baseline Systems}
We evaluate \sys{} against three fixed-size buffers spanning our implementation's capacitance range---770 $\mu{}$F, 10 mF, and 18 mF---to ensure the realized improvement is a result of energy-adaptive behavior rather than simply different buffer capacity.
To compare \sys{}'s capacitor architecture to prior work on dynamic energy buffers, we also implement and evaluate Morphy~\cite{morphy} for a similar capacitance range.
Our Morphy implementation uses eight 2 $mF$ capacitors with leakage current of approximately 25.2 $\mu{}A$ at 6.3V~\cite{myElectrolytics} (\ie{} slightly lower leakage than the capacitors in \sys{}).

Morphy uses a secondary microcontroller powered by a battery or backup capacitor to control the capacitor array; we use a second MSP430FR5994 powered through USB, corresponding to Morphy's battery-powered design.
Accordingly, we expect our results to slightly overestimate Morphy's performance in the fully-batteryless case as the system does not have to power the Morphy controller or charge a backup capacitor in our implementation.
Seven of the eight capacitors in the array are available to reconfigure, with one task capacitor kept in parallel to smooth voltage fluctuations from switching.
We evaluate the same subset of eleven possible configurations for the remaining seven capacitors as is done in the original Morphy work, resulting in a capacitance range for our Morphy implementation of 250 $\mu{}F$-16 $mF$.

\subsection{Computational Backend}
To explore how \sys{} affects performance across a range of system demands---focusing on diverse reactivity and longevity requirements---we implement four software benchmarks:

\begin{itemize}
    \item \textbf{Data Encryption (DE)}: Continuously perform AES-128 encryptions in software. This application has no reactivity requirements, low persistence requirements, and a predictable power draw; we use it as a baseline to explore \sys{}'s software and power overhead.
    \item \textbf{Sense and Compute (SC)}: Exit a deep-sleep mode once every five seconds second to sample and digitally filter readings from a low-power microphone~\cite{microphone}. This benchmark represents systems which value high reactivity and can accept low persistence; individual atomic measurements are low-energy, but the system must be online to take the measurements.
    \item \textbf{Radio Transmission (RT)}: Send buffered data over radio~\cite{ZL7051, wakeup} to a base station. Data transmission is an example of an application with high persistence requirements (radio transmissions are atomic and energy-intensive) and low reactivity requirements (transmitting data may be delayed until energy is available).
    \item \textbf{Packet Forwarding (PF)}: Listen for and retransmit unpredictable incoming data over the radio. Timely packet forwarding demands both high persistence and reactivity to successfully receive and retransmit data.
\end{itemize}

We emulate the power consumption of the necessary peripherals for each benchmark by toggling a resistor connected to a digital output on the MSP430, with values for each benchmark chosen to match the relevant peripheral.
The reactivity-focused benchmarks (SC and PF) have deadlines that may arrive while the system is off; we use a secondary MSP430 to deliver these events.
A deployed system may use remanence-based timekeepers~\cite{botoks} to track internal deadlines despite power failures for the SC benchmark, while incoming packets as in the PF benchmark would arrive from other systems.
Although we evaluate each benchmark in isolation, full systems are likely to exercise combinations of each requirement---one platform should support all reactivity, persistence, and efficiency requirements.

\subsection{Energy Harvesting Frontend}
\label{sec:eh-replay}
Energy harvesting volatility makes repeatable experimentation with batteryless devices difficult; uncontrollable environmental changes often have an outsized effect on energy input and obfuscate differences in actual system performance.
We make our experiments repeatable and consistent using a programmable power frontend inspired by the Ekho~\cite{ekho} record-and-replay platform.
The power controller supplies the energy buffer using a high-drive Digital-to-Analog Converter (DAC), measures the load voltage and input current using a sense resistor, and tunes the DAC to supply a programmed power level.
We evaluate \sys{} emulating both solar (5 $cm^2$, 22\% efficient cell~\cite{mini-solar-panel}) and RF energy (915 MHz dipole antenna~\cite{powercast-dipole}).
We also emulate the load-dependent performance of a commercial RF-to-DC converter~\cite{p2110b} and solar panel management chip~\cite{bq25570}.

\section{Evaluation}
\label{sec:evaluation}
\begin{table*}[t]
\setlength{\tabcolsep}{4.5pt}
\resizebox{\textwidth}{!}{
\centering
\footnotesize{}
\begin{tabular}{c|ccccc|ccccc|ccccc}
                    & \multicolumn{5}{c|}{\textbf{Data Encrypt}}                                     & \multicolumn{5}{c|}{\textbf{Sense and Compute}}                                & \multicolumn{5}{c}{\textbf{Radio Transmit}}                                    \\
\textbf{Buffer}     & \textbf{770$\mu{}$} & \textbf{10m} & \textbf{17m} & \textbf{Morphy} & \textbf{REACT} & \textbf{770$\mu{}$} & \textbf{10m} & \textbf{17m} & \textbf{Morphy} & \textbf{REACT} & \textbf{770$\mu{}$} & \textbf{10m} & \textbf{17m} & \textbf{Morphy} & \textbf{REACT} \\ \hline
\textbf{RF Cart}    & 1275          & 1574         & 1831         & 1745            & 1711           & 50            & 81           & 104          & 77              & 83             & 22            & 53           & 56           & 38              & 48             \\
\textbf{RF Obs.}    & 666           & 472          & 0            & 357             & 576            & 44            & 28           & 0            & 39              & 49             & 4             & 6            & 0            & 0               & 3              \\
\textbf{RF Mob.}    & 810           & 1004         & 645          & 801             & 1038           & 52            & 50           & 40           & 53              & 84             & 4             & 13           & 12           & 4               & 15             \\
\textbf{Sol. Camp.} & 6666          & 7290         & 7936         & 8194            & 9756           & 330           & 353          & 367          & 398             & 439            & 1376          & 1457         & 1542         & 1059            & 1426           \\
\textbf{Sol. Comm.} & 2168          & 2186         & 2554         & 2399            & 2232           & 88            & 110          & 130          & 133             & 154            & 8             & 40           & 48           & 31              & 34             \\ \hline
\textbf{Mean}       & 2317          & 2505         & 2593         & 2699            & 3063           & 113           & 124          & 128          & 140             & 162            & 283           & 314          & 332          & 226             & 313           
\end{tabular}
}
\caption{Performance on the DE, SC, and RT benchmarks, across traces and energy buffers.}
\label{table:encrypt-sense-transmit}
\hrulefill
\end{table*}
\begin{table}[t]
\centering
\resizebox{\columnwidth}{!}{
\footnotesize{}
%
\begin{tabular}{c|ccc}
\textbf{Trace}          & \textbf{Time (s)} & \textbf{Avg. Pow. (mW)} & \textbf{Power CV*} \\ \hline
\textbf{RF Cart}        & 313               & 2.12                    & 103\%              \\
\textbf{RF Obstruction} & 313               & 0.227                   & 61\%               \\
\textbf{RF Mobile}      & 318               & 0.5                     & 166\%              \\ \hline
\textbf{Solar Campus}   & 3609              & 5.18                    & 207\%              \\
\textbf{Solar Commute}  & 6030              & 0.148                   & 333\%             
\end{tabular}
}
\caption{Details of each power trace. *CV = Coefficient of Variation.}
\label{table:traces}
\hrulefill
\end{table}
We evaluate \sys{} alongside the baseline buffers running each benchmark under three RF and two solar traces from publicly available repositories~\cite{enhants, rf-traces-anon}, representative of power dynamics for small energy harvesting systems.
We record the RF traces in an active office environment using a commercial harvester and transmitter~\cite{p2110b, tx91501b} and use solar irradiance traces from the Enhants mobile irradiance dataset~\cite{enhants}; Table~\ref{table:traces} gives a short summary of each trace.
These traces show the power variability common for IoT-scale harvesters: environmental changes (\eg{} ambient RF levels, time of day) affect \textit{average} input power, while short-term changes such as orientation cause instantaneous variation even if the environment is unchanged.
We apply each trace using the power replay system described in \S{}~\ref{sec:eh-replay}; once the trace is complete, we let the system run until it drains the buffer capacitor.

\subsection{Characterization and Overhead}
Figure~\ref{fig:scope-comparison} illustrates \sys{}'s behavior through the last-level buffer voltage when varying capacitance; the inset focuses on \sys{}'s voltage output as it expands to capture energy (also shown is the voltage of the comparable Morphy array).
From a cold start \sys{} only charges the last-level buffer---rapidly reaching the enable voltage and then the upper voltage threshold (3.5V).
\sys{} then adds a series-configured capacitor bank to capture excess incoming energy.
Voltage drops as the system temporarily operates exclusively from the last-level buffer while harvested energy goes towards charging the new capacitance.
As power input falls, \sys{}'s output voltage falls below the upper threshold voltage---indicating \sys{} is operating at an efficient capacitance point.
At $t\approx{}450s$ the last-level buffer is discharged to the lower threshold and \sys{} begins switching banks into series mode to boost their output voltage and charge the last-level buffer, visible in Figure~\ref{fig:scope-comparison} as five voltage spikes corresponding to each capacitor bank---sustaining operation until no more energy is available at $t\approx{}500s$.

We characterize \sys{}'s software overhead by running the DE benchmark on continuous power for 5 minutes with and without \sys{}'s software component, which periodically interrupts execution to measure the capacitor bank.
At a sample rate of 10 Hz, \sys{} adds a 1.8\% penalty to software-heavy applications.
We measure \sys{}'s power overhead by comparing the execution time of systems running the DE benchmark using \sys{} and the 770 $\mu{}F$ buffer after charging each to their enable voltage.
Based on this approach we estimate that our implementation of \sys{} introduces a 68 $\mu{}W$ power draw, or \mytilde{}14 $\mu{}W$ per bank.

\subsection{\sys{} Minimizes System Latency}
\begin{table}[t]
\centering
\resizebox{\columnwidth}{!}{
\footnotesize{}
\begin{tabular}{c|ccccc}
\textbf{Buffer}     & \textbf{770 $\mu{}F$} & \textbf{10 mF} & \textbf{17 mF} & \textbf{Morphy} & \textbf{REACT} \\ \hline
\textbf{RF Cart}    & 6.65            & 17.73          & 31.27          & 5.51            & 6.65           \\
\textbf{RF Obs.}    & 14.58           & 223.07         & -              & 6.50            & 16             \\
\textbf{RF Mob.}    & 6.90            & 148.10         & 239.88         & 5.65            & 6.38           \\
\textbf{Sol. Camp.} & 42.11           & 737.39         & 741.42         & 35.59           & 41.26          \\
\textbf{Sol. Comm.} & 119.60          & 196.30         & 213.00         & 108.10          & 130.6          \\ \hline
\textbf{Mean}       & 37.97           & 264.92         & 306.39         & 32.27           & 40.18         
\end{tabular}
}
\caption{System latency (seconds) across traces and energy buffers. - indicates system never begins operation.}
\label{table:charge}
\hrulefill
\end{table}
Table~\ref{table:charge} details the time it takes each system to begin operation, across power traces and energy buffers (charge time is software-invariant and constant across benchmarks).
Latency is driven by both capacitor size and environment---the $10 mF$ buffer is \mytilde{}$13x$ larger than the $770 \mu{}F$ buffer and takes on average $7x$ longer to activate the system across our traces.
High-capacity static buffers incur a larger latency penalty even if mean power input is high if much of that power is contained in a short-term spike later in the trace (\eg{} for the Solar Campus trace), but these dynamics are generally impossible to predict at design time.
By exclusively charging the last-level buffer while the rest of the system is off, \sys{} matches the latency of the smallest static buffer---an average of $7.7x$ faster than the equivalent-capacity 17 mF buffer, which risks failing to start at all.
Morphy further reduces system latency because its smallest configuration is smaller than \sys{}'s last-level buffer (250 $\mu{}F$ vs 770 $\mu{}F$), although the limited reduction in average latency compared to the reduction in capacitance (Morphy realizes an average 20\% reduction in latency over \sys{} using a 68\% smaller capacitance) suggests that further reducing capacitance yields diminishing latency returns in realistic energy environments.

Minimizing latency improves reactivity-bound applications such as the SC and PF benchmarks; this effect is visible in Table~\ref{table:encrypt-sense-transmit} as the 770 $\mu{}F$ buffer outperforms larger static versions in the SC benchmark for relatively low-power traces (RF Mobile/Obstructed).
\sys{} inherits the latency advantage due to the small last-level buffer, similarly improving performance on each power trace.
Morphy realizes a similar performance improvement over the static systems, but ultimately underperforms \sys{} as a result of inefficient capacitor switching (\S{}~\ref{sec:performance}).
Small static buffers enable low-latency operation, but at the cost of energy capacity.
As power input increases, the latency penalty of large buffers fades and their increased capacity enables them to operate for longer---resulting in higher performance for larger static buffers under high-power traces (RF Cart, Solar Campus).
Smaller buffers, in turn, become less efficient as they must burn more incoming energy off as waste heat.

\subsection{\sys{} Maximizes Energy Capacity}
\begin{figure}[]
  \centering
  \includegraphics[width=\columnwidth]{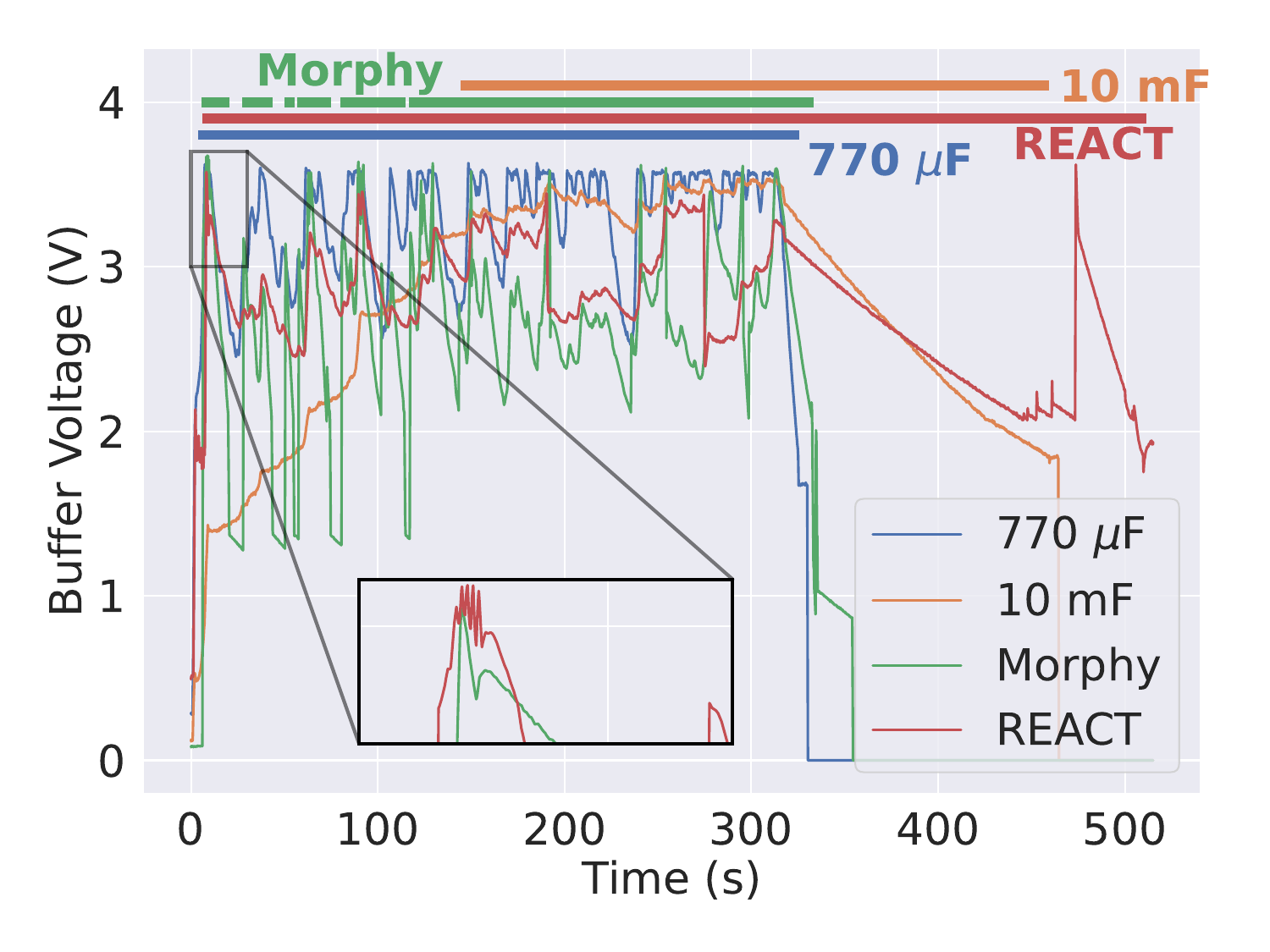}
  \caption{Buffer voltage and on-time for the SC benchmark under RF Mobile power. Solid bars indicate when the system is operating.}
  \label{fig:scope-comparison}
  \hrulefill
\end{figure}
Figure~\ref{fig:scope-comparison} illustrates the system-level effects of the capacity-latency tradeoff, and how \sys{} avoids this tradeoff through energy-adaptive buffering.
The small $770 \mu{}F$ buffer charges rapidly, but reaches capacity and discharges energy when it does not have work to match incoming power (illustrated by clipping at 3.6V on the 770 $\mu{}$F line).
The 10 mF buffer sacrifices latency for capacity---starting operation 21x later than the smaller buffer, but avoiding overvoltage.
Morphy begins execution early with a small capacitance, but its lossy switching mechanism means it does not overall outperform the $770 \mu{}F$ buffer.
In contrast, \sys{} achieves low latency, high efficiency, \textit{and} high capacity by efficiently expanding capacitance as necessary after enabling the system.

Tables~\ref{table:encrypt-sense-transmit} and \ref{table:forwarding} show that high capacity is valuable when average input power exceeds output power (\eg{} DE and SC benchmarks executed under the RF Cart trace), or when peak power demand is uncontrollable and uncorrelated with input (\eg{} the PF benchmark executed on Solar Campus, where both power supply and demand are concentrated in short bursts).
In both cases, high-capacity systems store excess energy to continue operation even if future power input falls or demand rises.
\sys{} efficiently expands to capture all incoming energy during periods of high net input power, matching or beating the performance of the 10 mF and 17 mF systems when they outperform the small 770 $\mu{}F$ buffer.

\subsection{\sys{} Provides Flexible, Efficient Longevity}
\label{sec:longevity}
\begin{table}[t]
\setlength{\tabcolsep}{2.9pt}
\centering
\resizebox{\columnwidth}{!}{
\footnotesize
\begin{tabular}{c|cc|cc|cc|cc|cc}
\textbf{Buffer}     & \multicolumn{2}{c|}{\textbf{770 uF}} & \multicolumn{2}{c|}{\textbf{10 mF}} & \multicolumn{2}{c|}{\textbf{17 mF}} & \multicolumn{2}{c|}{\textbf{Morphy}} & \multicolumn{2}{c}{\textbf{REACT}} \\ \hline
\textbf{Packets}    & Rx                & Tx               & Rx               & Tx               & Rx               & Tx               & Rx                & Tx               & Rx               & Tx              \\ \hline
\textbf{RF Cart}    & 22                & 10               & 49               & 49               & 48               & 48               & 55                & 22               & 53               & 52              \\
\textbf{RF Obs.}    & 4                 & 4                & 4                & 4                & 0                & 0                & 2                 & 0                & 3                & 0               \\
\textbf{RF Mob.}    & 11                & 4                & 14               & 13               & 9                & 9                & 19                & 0                & 38               & 5               \\
\textbf{Sol. Camp.} & 163               & 163              & 240              & 240              & 196              & 196              & 206               & 204              & 284              & 277             \\
\textbf{Sol. Comm.} & 72                & 8                & 35               & 35               & 33               & 33               & 85                & 14               & 84               & 63              \\ \hline
\textbf{Mean}       & 54                & 38               & 68               & 68               & 57               & 57               & 73                & 48               & 92               & 80             
\end{tabular}
}
\caption{Packets successfully received and retransmitted during the Packet Forwarding benchmark.}
\label{table:forwarding}
\hrulefill
\end{table}
We evaluate \sys{}'s software-directed longevity guarantees (\S{}~\ref{sec:software-longevity}) on the longevity-bound RT and PF benchmarks.
We compare \sys{} to the 770 $\mu{}F$ buffer, which cannot sustain a full transmission without additional input power.
Running the RT benchmark under the RF Cart trace isolates this limitation as the 770 $\mu{}F$ static buffer significantly underperforms the other buffers despite never reaching capacity: instead, it wastes power on doomed-to-fail transmissions when incoming power cannot make up for the deficit.
We augment the RT benchmark code for our \sys{} implementation to include a minimum capacitance level for \sys{}, below which the system waits to gather more energy in a low-power sleep mode.
Leveraging \sys{}'s variable capacitance allows software to buffer energy to guarantee completion, more than doubling the number of successful transmissions and ultimately outperforming even the larger buffers.

We use the same approach to execute the RT benchmark on our Morphy implementation.
Similar to \sys{}, Morphy varies capacitance to keep supply voltage within an acceptable level for the application microcontroller while also waiting to gather enough energy to power a full transmission.
Morphy's underperformance compared to both \sys{} and the static buffers is a result of Morphy's capacitor network design---as Morphy reconfigures the capacitor array to increase capacitance, stored energy is dissipated as current flows between capacitors in the network.
This energy dissipation dramatically reduces Morphy's end-to-end performance, particularly in systems where Morphy \textit{must} switch capacitance to ensure success (\ie{} the RT and PF benchmarks).
\sys{}'s isolated capacitor banks eliminate this problem by restricting current flow during switching; the energy savings are reflected in the end-to-end performance, where \sys{} completes on average 38\% more transmissions than Morphy.

\subsubsection{Fungible Energy Storage}
\label{sec:fungibility}
A unified buffer means that energy is fungible, and \sys{} is flexible: software can re-define or ignore previous longevity requirements if conditions change or a higher-priority task arrives.
The PF benchmark (Table~\ref{table:forwarding}) shows the value of energy fungibility using two tasks with distinct reactivity and longevity requirements.
Receiving an incoming packet requires a moderate level of longevity, but is uncontrollable and has a strict reactivity requirement (the system can only receive a packet exactly when it arrives).
Re-transmission requires more energy and thus more longevity, but has no deadline.
Software must effectively split energy between a controllable high-power task and an uncontrollable lower-power task.

As in the RT benchmark we use the minimum-capacitance approach to set separate longevity levels for each task, using a similar approach for our Morphy implementation.
When the system has no packets to transmit, it waits in a deep-sleep until receiving an incoming packet.
If \sys{} contains sufficient energy when the packet arrives, it receives and buffers the packet to later send.
\sys{} then begins charging for the transmit task, forwarding the buffered packet once enough energy is available.
If another packet is received while \sys{} is charging for the transmit task, however, software disregards the transmit-associated longevity requirement to execute the receive task if sufficient energy is available.

Table~\ref{table:forwarding} shows that \sys{} outperforms all static buffer designs on the PF benchmark by efficiently addressing the requirements of both tasks, resulting in a mean performance improvement of 54\%.
\sys{}'s maximal reactivity enables it to turn on earlier and begin receiving and buffering packets to send during later periods of high power, while its high capacity enables it to make the most of incoming energy during those high periods.
Software-level longevity guarantees both ensure the system only begins receive/transmit operations when enough energy is available to complete them, and that software can effectively allocate energy to incoming events as needed.
Although Morphy enables the same software-level control of energy allocation, the energy dissipated when switching capacitors in the interconnected array means that Morphy's overall performance on the PF benchmark is below that of the best performing static buffer.

\subsection{\sys{} Improves End-to-End System Efficiency}
\label{sec:performance}
\begin{figure}[t]
  \centering
  \includegraphics[width=0.95\columnwidth]{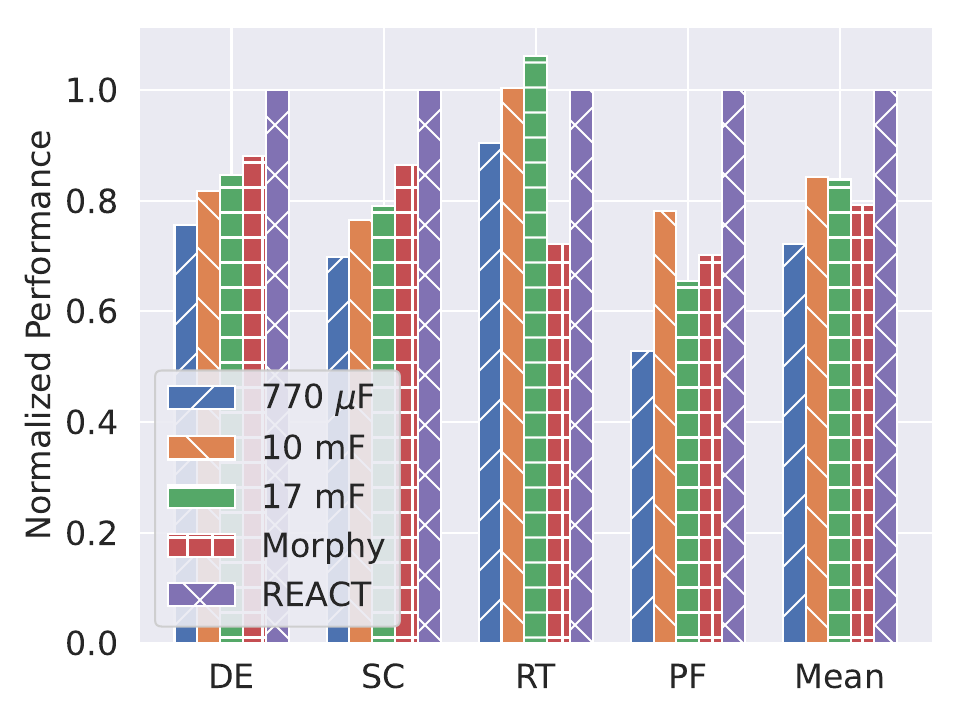}
  \caption{Average buffer performance quantified by figures of merit across power traces for each benchmark, normalized to \sys{}.}
  \label{fig:performance-bars}
  \hrulefill
\end{figure}
Optimizing buffer behavior maximizes the amount of harvested energy available to the end system for useful work.
Figure~\ref{fig:performance-bars} illustrates the aggregate performance of \sys{} compared to the baseline systems across the benchmarks and power traces we evaluate; we find that \sys{} improves performance over the equally-reactive 770 $\mu{}F$ buffer by an average of 39.1\%, over the equal-capacity 17 mF buffer by 19.3\%, and over the next-best-efficient 10 mF buffer by 18.8\%.
Compared to Morphy, \sys{} improves aggregate performance by 26.2\%---demonstrating the necessity of \sys{}'s bank isolation approach and boosting performance where prior dynamic capacitance systems underperform static approaches.
Extreme cases where the system is \textit{always} operating in an energy surplus or deficit---such as the low-power SC benchmark under the high-power RF Cart trace---the extra power consumption from \sys{}'s hardware causes it to underperform suitable static buffers because \sys{}'s flexibility is unnecessary.
In the common case, however, volatile power conditions expose the latency, longevity, and efficiency-related shortcomings of static buffer designs and expose the value of \sys{}'s efficient variable-capacitance approach.

\section{Conclusion}
\label{sec:conclusion}
Energy harvesting systems operate on unreliable and volatile power, but use fixed-size buffers which waste energy and functionally limit systems when allocated capacity is a poor fit for short-term power dynamics.
\sys{} stores incoming energy in a fabric of reconfigurable capacitor banks, varying equivalent capacitance according to current energy supply and demand dynamics---adding capacitance to capture surplus power and reclaiming energy from excess capacitance.
\sys{}'s energy-adaptive approach maximizes reactivity and capacity to ensure all incoming energy is captured and efficiently delivered to sensing, computing, and communication devices.
Our hardware evaluation on real-world power traces shows that \sys{} reduces system latency by an average of 7.7x compared to an equivalent-sized static buffer and improves throughput by an average of 25.6\% over \textit{any} static buffer system, while incorporating software direction allows \sys{} to provide flexible and fungible task longevity guarantees.
Compared to state-of-the-art switched capacitor systems, \sys{}'s efficient switching architecture improves performance by an average of 26.2\%.

\sys{}'s runtime-configurable buffering technique eliminates the tradeoff between system latency and longevity, and affords designers greater control over how batteryless devices respond to incoming power.
Our results indicate that energy-responsive reconfiguration of hardware is an effective approach to both maximizing energy efficiency and system functionality, opening the door for future work leveraging energy-adaptive hardware and reconfiguration.

\bibliographystyle{plain}
\bibliography{refs}

\begin{thebibliography}{10}

\bibitem{d2vfs}
Saad Ahmed, Qurat~ul Ain, Junaid~Haroon Siddiqui, Luca Mottola, and
  Muhammad~Hamad Alizai.
\newblock Intermittent computing with dynamic voltage and frequency scaling.
\newblock In {\em Proceedings of the 2020 International Conference on Embedded
  Wireless Systems and Networks}, EWSN '20, page 97–107, USA, 2020. Junction
  Publishing.

\bibitem{concrete}
Miran Alhaideri, Michael Rushanan, Denis~Foo Kune, and Kevin Fu.
\newblock The moo and cement shoes: Future directions of a practical
  sense-control-actuate application, September 2013.
\newblock Presented at First International Workshop on the Swarm at the Edge of
  the Cloud (SEC'13 @ ESWeek), Montreal.

\bibitem{rf-traces-anon}
Anon.
\newblock Rf traces, October 2022.
\newblock \url{https://anonymous.4open.science/r/rf_traces-4B3E/README.md}.

\bibitem{rehash}
Abu Bakar, Alexander~G. Ross, Kasim~Sinan Yildirim, and Josiah Hester.
\newblock Rehash: A flexible, developer focused, heuristic adaptation platform
  for intermittently powered computing.
\newblock {\em Proc. ACM Interact. Mob. Wearable Ubiquitous Technol.}, 5(3),
  sep 2021.

\bibitem{what-is-mmtc}
James Blackman.
\newblock What is mmtc in 5g nr, and how does it impact nb-iot and lte-m,
  October 2019.
\newblock
  https://enterpriseiotinsights.com/20191016/channels/fundamentals/what-is-mmtc-in-5g-nr-and-how-does-it-impact-nb-iot-and-lte-m.

\bibitem{dewdrop}
Michael Buettner, Ben Greenstein, and David Wetherall.
\newblock Dewdrop: An energy-aware runtime for computational rfid.
\newblock In {\em Proceedings of the 8th USENIX Conference on Networked Systems
  Design and Implementation}, NSDI’11, page 197–210, USA, 2011. USENIX
  Association.

\bibitem{capybara}
Alexei Colin, Emily Ruppel, and Brandon Lucia.
\newblock A reconfigurable energy storage architecture for energy-harvesting
  devices.
\newblock In {\em Proceedings of the Twenty-Third International Conference on
  Architectural Support for Programming Languages and Operating Systems},
  ASPLOS '18, page 767–781, New York, NY, USA, 2018. Association for
  Computing Machinery.

\bibitem{botoks}
Jasper de~Winkel, Carlo Delle~Donne, Kasim~Sinan Yildirim, Przemys\l{}aw
  Pawe\l{}czak, and Josiah Hester.
\newblock Reliable timekeeping for intermittent computing.
\newblock In {\em Proceedings of the Twenty-Fifth International Conference on
  Architectural Support for Programming Languages and Operating Systems},
  ASPLOS '20, page 53–67, New York, NY, USA, 2020. Association for Computing
  Machinery.

\bibitem{gameboy}
Jasper de~Winkel, Vito Kortbeek, Josiah Hester, and Przemys\l{}aw Pawe\l{}czak.
\newblock Battery-free game boy.
\newblock {\em Proc. ACM Interact. Mob. Wearable Ubiquitous Technol.}, 4(3),
  sep 2020.

\bibitem{phase}
H.~{Desai} and B.~{Lucia}.
\newblock A power-aware heterogeneous architecture scaling model for
  energy-harvesting computers.
\newblock {\em IEEE Computer Architecture Letters}, 19(1):68--71, 2020.

\bibitem{microphone}
Knowles Electronics.
\newblock {SPU0414HR5H-SB}, December 2012.
\newblock
  \url{https://www.mouser.com/datasheet/2/218/knowles_01232019_SPU0414HR5H_SB-1891952.pdf}.

\bibitem{enhants}
M.~{Gorlatova}, A.~{Wallwater}, and G.~{Zussman}.
\newblock Networking low-power energy harvesting devices: Measurements and
  algorithms.
\newblock In {\em 2011 Proceedings IEEE INFOCOM}, pages 1602--1610, 2011.

\bibitem{flowmeter}
Wang~Song Hao and Ronald Garcia.
\newblock Development of a digital and battery-free smart flowmeter.
\newblock {\em Energies}, 7(6):3695--3709, 2014.

\bibitem{ekho}
Josiah Hester, Timothy Scott, and Jacob Sorber.
\newblock Ekho: Realistic and repeatable experimentation for tiny
  energy-harvesting sensors.
\newblock In {\em Proceedings of the 12th ACM Conference on Embedded Network
  Sensor Systems}, SenSys '14, page 330–331, New York, NY, USA, 2014.
  Association for Computing Machinery.

\bibitem{federation}
Josiah Hester, Lanny Sitanayah, and Jacob Sorber.
\newblock Tragedy of the coulombs: Federating energy storage for tiny,
  intermittently-powered sensors.
\newblock In {\em ACM Conference on Embedded Networked Sensor Systems}, SenSys,
  pages 5--16, 2015.

\bibitem{flicker}
Josiah Hester and Jacob Sorber.
\newblock Flicker: Rapid prototyping for the batteryless internet-of-things.
\newblock In {\em Proceedings of the 15th ACM Conference on Embedded Network
  Sensor Systems}, SenSys '17, New York, NY, USA, 2017. Association for
  Computing Machinery.

\bibitem{clank}
Matthew Hicks.
\newblock Clank: Architectural support for intermittent computation.
\newblock In {\em International Symposium on Computer Architecture}, {ISCA},
  pages 228--240, 2017.

\bibitem{wakeup}
Fraunhofer IIS.
\newblock {RFicient Basic, Ultra-Low-Power WakeUp Receiver}, January 2019.
\newblock
  \url{https://www.iis.fraunhofer.de/content/dam/iis/en/doc/il/ics/ic-design/Datenblaetter/Factsheet_WakeUp_v4.pdf}.

\bibitem{msp430g2252}
Texas Instruments.
\newblock Msp430g2x52, msp430g2x12 mixed signal microcontroller datasheet (rev.
  g), May 2013.
\newblock \url{https://www.ti.com/lit/ds/symlink/msp430g2252.pdf}.

\bibitem{bq25570}
Texas Instruments.
\newblock {bq25570 nano power boost charger and buck converter for energy
  harvester powered applications}, March 2019.
\newblock \url{https://www.ti.com/lit/ds/symlink/bq25570.pdf}.

\bibitem{lm66100}
Texas Instruments.
\newblock Lm66100 5.5-v, 1.5-a 79-milliohm, low iq ideal diode with input
  polarity protection, June 2019.
\newblock \url{https://www.ti.com/lit/ds/symlink/lm66100.pdf}.

\bibitem{fr5994}
Texas Instruments.
\newblock {MSP430FR599x, MSP430FR596x Mixed-Signal Microcontrollers}, January
  2021.
\newblock \url{https://www.ti.com/lit/ds/symlink/msp430fr5994.pdf}.

\bibitem{dust}
Joseph Kahn, Randy Katz, and Kristofer Pister.
\newblock {Next Century Challenges: Mobile Networking for ''Smart Dust''}.
\newblock In {\em Conference on Mobile Computing and Networking (MobiCom)},
  1999.

\bibitem{supercap}
Kemet.
\newblock Supercapacitors fm series, July 2020.
\newblock
  \url{https://www.mouser.com/datasheet/2/212/1/KEM_S6012_FM-1103835.pdf}.

\bibitem{efm32wg}
Silicon Labs.
\newblock {EFM32 Gecko Family EFM32WG Data Sheet}, December 2021.
\newblock
  \url{https://www.silabs.com/documents/public/data-sheets/efm32wg-datasheet.pdf}.

\bibitem{dino}
Brandon Lucia and Benjamin Ransford.
\newblock A simpler, safer programming and execution model for intermittent
  systems.
\newblock In {\em Conference on {Programming} {Language} {Design} and
  {Implementation}}, {PLDI}, pages 575--585, 2015.

\bibitem{nvprocessor}
K.~{Ma}, Y.~{Zheng}, S.~{Li}, K.~{Swaminathan}, X.~{Li}, Y.~{Liu},
  J.~{Sampson}, Y.~{Xie}, and V.~{Narayanan}.
\newblock Architecture exploration for ambient energy harvesting nonvolatile
  processors.
\newblock In {\em IEEE International Symposium on High Performance Computer
  Architecture}, HPCA, pages 526--537, Feb 2015.

\bibitem{alpaca}
Kiwan Maeng, Alexei Colin, and Brandon Lucia.
\newblock Alpaca: Intermittent execution without checkpoints.
\newblock In {\em International {Conference} on {Object}-{Oriented}
  {Programming}, {Systems}, {Languages}, and {Applications}}, {OOPSLA}, pages
  96:1--96:30, October 2017.

\bibitem{chinchilla}
Kiwan Maeng and Brandon Lucia.
\newblock Adaptive dynamic checkpointing for safe efficient intermittent
  computing.
\newblock In {\em {USENIX} {Conference} on {Operating} {Systems} {Design} and
  {Implementation}}, {OSDI}, pages 129--144, November 2018.

\bibitem{catnap}
Kiwan Maeng and Brandon Lucia.
\newblock Adaptive low-overhead scheduling for periodic and reactive
  intermittent execution.
\newblock In {\em Proceedings of the 41st ACM SIGPLAN Conference on Programming
  Language Design and Implementation}, PLDI 2020, page 1005–1021, New York,
  NY, USA, 2020. Association for Computing Machinery.

\bibitem{ZL7051}
Microsemi.
\newblock {ZL70251 Ultra-Low-Power Sub-GHz RF Transceiver}, March 2018.
\newblock
  \url{https://www.microsemi.com/document-portal/doc_view/132900-zl70251-datasheet}.

\bibitem{myCeramics}
Murata.
\newblock {GRM31CR60J227ME11L Chip Monolithic Ceramic Capacitor for General}.
\newblock
  \url{https://search.murata.co.jp/Ceramy/image/img/A01X/G101/ENG/GRM31CR60J227ME11-01.pdf}.

\bibitem{mySupercap}
Murata.
\newblock {Supercapacitors FM Series}, July 2020.
\newblock
  \url{https://www.mouser.com/datasheet/2/212/1/KEM_S6012_FM-1103835.pdf}.

\bibitem{batteryless-ingestible}
Phillip Nadeua, Dina El-Damaj, Deal Glettig, Yong~Lin Kong, Stacy Mo, Cody
  Cleveland, Lucas Booth, Niclas Roxhed, Robert Langer, Anantha~P.
  Chandrakasan, and Giovanni Traverso.
\newblock Prolonged energy harvesting for ingestible devices.
\newblock {\em Nature Biomedical Engineering}, 1(0022), Feb 2017.

\bibitem{myElectrolytics}
Nichicon.
\newblock {ALUMINUM ELECTROLYTIC CAPACITORS}.
\newblock \url{https://www.nichicon.co.jp/english/products/pdfs/e-kl.pdf}.

\bibitem{cr2032}
Panasonic.
\newblock Panasonic coin type lithium batteries, August 2005.
\newblock
  \url{https://datasheet.octopart.com/CR1616-Panasonic-datasheet-9751741.pdf}.

\bibitem{p2110b}
Powercast.
\newblock {P2110B 915 MHz RF Powerharvester Receiver}, December 2016.
\newblock
  \url{https://www.powercastco.com/wp-content/uploads/2016/12/P2110B-Datasheet-Rev-3.pdf}.

\bibitem{tx91501b}
Powercast.
\newblock {TX91501B – 915 MHz Powercaster Transmitter}, October 2019.
\newblock
  \url{https://www.powercastco.com/wp-content/uploads/2019/10/User-Manual-TX-915-01B-Rev-A-1.pdf}.

\bibitem{powercast-dipole}
Powercast.
\newblock 915 mhz dipole antenna datasheet, November 2020.
\newblock
  \url{https://www.powercastco.com/wp-content/uploads/2020/11/DA-915-01-Antenna-Datasheet_new_web.pdf}.

\bibitem{mementos}
Benjamin Ransford, Jacob Sorber, and Kevin Fu.
\newblock {Mementos: System Support for Long-Running Computation on RFID-Scale
  Devices}.
\newblock In {\em Architectural Support for Programming Languages and Operating
  Systems (ASPLOS)}, 2011.

\bibitem{piezo}
Henry Sodano, Gyuhae Park, and Daniel Inman.
\newblock {Estimation of Electric Charge Output for Piezoelectric Energy
  Harvesting}.
\newblock In {\em Strain, Volume 40}, 2004.

\bibitem{schottky}
ST.
\newblock Small signal schottky diode, October 2001.
\newblock
  \url{https://www.st.com/content/ccc/resource/technical/document/datasheet/group1/11/76/e4/a3/df/07/49/14/CD00000767/files/CD00000767.pdf/jcr:content/translations/en.CD00000767.pdf}.

\bibitem{mini-solar-panel}
Voltaic.
\newblock Voltaic systems p121 r1g, April 2020.
\newblock \url{https://voltaicsystems.com/content/Voltaic Systems P121
  R1G.pdf}.

\bibitem{totalrecall}
Harrison Williams, Xun Jian, and Matthew Hicks.
\newblock Forget failure: Exploiting sram data remanence for low-overhead
  intermittent computation.
\newblock In {\em Proceedings of the Twenty-Fifth International Conference on
  Architectural Support for Programming Languages and Operating Systems},
  ASPLOS '20, page 69–84, New York, NY, USA, 2020. Association for Computing
  Machinery.

\bibitem{failureSentinels}
Harrison Williams, Michael Moukarzel, and Matthew Hicks.
\newblock Failure sentinels: Ubiquitous just-in-time intermittent computation
  via low-cost hardware support for voltage monitoring.
\newblock In {\em International Symposium on Computer Architecture}, {ISCA},
  pages 665--678, 2021.

\bibitem{ratchet}
Joel Van~Der Woude and Matthew Hicks.
\newblock Intermittent computation without hardware support or programmer
  intervention.
\newblock In {\em {USENIX} {Symposium} on {Operating} {Systems} {Design} and
  {Implementation}}, {OSDI}, pages 17--32, November 2016.

\bibitem{rice-computer}
X.~{Wu}, I.~{Lee}, Q.~{Dong}, K.~{Yang}, D.~{Kim}, J.~{Wang}, Y.~{Peng},
  Y.~{Zhang}, M.~{Saliganc}, M.~{Yasuda}, K.~{Kumeno}, F.~{Ohno}, S.~{Miyoshi},
  M.~{Kawaminami}, D.~{Sylvester}, and D.~{Blaauw}.
\newblock A 0.04mm316nw wireless and batteryless sensor system with integrated
  cortex-m0+ processor and optical communication for cellular temperature
  measurement.
\newblock In {\em 2018 IEEE Symposium on VLSI Circuits}, pages 191--192, 2018.

\bibitem{astar}
Fan Yang, Ashok~Samraj Thangarajan, Wouter Joosen, Christophe Huygens, Danny
  Hughes, Gowri~Sankar Ramachandran, and Bhaskar Krishnamachari.
\newblock Astar: Sustainable battery free energy harvesting for heterogeneous
  platforms and dynamic environments.
\newblock In {\em Proceedings of the 2019 International Conference on Embedded
  Wireless Systems and Networks}, EWSN '19, page 71–82, USA, 2019. Junction
  Publishing.

\bibitem{morphy}
Fan Yang, Ashok~Samraj Thangarajan, Sam Michiels, Wouter Joosen, and Danny
  Hughes.
\newblock Morphy: Software defined charge storage for the iot.
\newblock In {\em Proceedings of the 19th ACM Conference on Embedded Networked
  Sensor Systems}, SenSys '21, page 248–260, New York, NY, USA, 2021.
  Association for Computing Machinery.

\end{thebibliography}


\end{document}